\documentclass[aip, amsmath, amssymb, reprint]{revtex4-1}

\usepackage{graphicx}
\usepackage{dcolumn}
\usepackage{bm}

\usepackage[T1]{fontenc}
\usepackage[utf8]{inputenc}
\usepackage{amsmath}
\usepackage{float}

\draft

\begin{document}

\title[Absolute Seebeck coefficient of thin platinum films]{Absolute Seebeck coefficient of thin platinum films}

\author{M. Kockert}
\email{kockert@physik.hu-berlin.de}
\affiliation{Novel Materials Group, Humboldt-Universität zu Berlin, 10099 Berlin, Germany}
\author{R. Mitdank}
\affiliation{Novel Materials Group, Humboldt-Universität zu Berlin, 10099 Berlin, Germany}
\author{A. Zykov}
\affiliation{Nanoscale processes and growth, Humboldt-Universität zu Berlin, 10099 Berlin, Germany}
\author{S. Kowarik}%
\affiliation{Nanoscale processes and growth, Humboldt-Universität zu Berlin, 10099 Berlin, Germany}
\affiliation{Bundesamt f\"{u}r Materialforschung und -pr\"{u}fung (BAM), 12203 Berlin, Germany}
\author{S. F. Fischer}
\email{saskia.fischer@physik.hu-berlin.de}
\affiliation{Novel Materials Group, Humboldt-Universität zu Berlin, 10099 Berlin, Germany}

\date{\today}

\begin{abstract}

The influence of size effects on the thermoelectric properties of thin platinum films is investigated and compared to the bulk. Structural properties, like the film thickness and the grain size, are varied. We correlate the electron mean free path with the temperature dependence of the electrical conductivity and the absolute Seebeck coefficient $S_{\text{Pt}}$ of platinum. We use a measurement platform as a standardized method to determine $S_{\text{Pt}}$ and show that $S_{\text{Pt,film}}$ is reduced compared to $S_{\text{Pt,bulk}}$. Boundary and surface scattering reduce the thermodiffusion and the phonon drag contribution to $S_{\text{Pt,film}}$ by nearly the same factor. A detailed discussion and a model to describe the temperature dependence of the absolute Seebeck coefficient and the influence of size effects of electron-phonon and phonon-phonon interaction on $S_{\text{Pt}}$ is given. 

\end{abstract}

\maketitle

\section{\label{sec:introduction}Introduction}

Platinum is the most commonly used thermoelectric reference material and is used with other materials, e.g. as commercially available bulk thermocouples \cite{Machin}. However, in recent years micro- and nanopatterning have become more interesting \cite{Burke,Dresselhaus}. Popular examples are thin films \cite{Venkatasubramanian,Daniel} and nanowires \cite{Kojda2,Hochbaum}. New challenges for metrology and its interpretation are coming with this trend. 

In order to determine the thermoelectric transport properties of nanowires, measurements are usually performed relative to thin films \cite{Kojda1,Kojda2,Hochbaum,KimJ}. For this purpose, microelectromechanical systems (MEMS) with thin platinum conduction lines of a few hundred nanometer thickness have been developed as measurement platforms \cite{Wang,Moosavi,Linseis}. However, thin metal films have a reduced absolute Seebeck coefficient $S_{\text{film}}$ compared to the bulk $S_{\text{bulk}}$ \cite{Brandli,HuebenerPt1,HuebenerAu1,LeonardCu,LeonardAg,Das}. Especially for metal-metal junctions, it is important to know the absolute Seebeck coefficient of the reference material. Deviations in the single-digit microvolt per Kelvin range can easily lead to misinterpretations of the measurement results. 

For this reason, we present a measurement platform to investigate the temperature-dependent thermoelectric transport properties of thin metal films. We demonstrate the platforms usability by investigating platinum films with a thickness of 134\,nm and 197\,nm.
To understand the influence of the microstructure on the absolute Seebeck coefficient $S$, we adapted and improved a model \cite{MacDonald,HuebenerPt2} that allows the decomposition of $S$ into a thermodiffusion and a phonon drag contribution. We discuss the influence of electron-phonon interaction compared to the phonon-phonon interaction on the phonon drag part of $S_{\text{Pt,film}}$ and $S_{\text{Pt,bulk}}$. This work shows that thin platinum films differ significantly from the bulk in terms of $S$ by nearly 400\,\% at $T=290$\,K.

\section{\label{sec:exp-details}Experimental Details}

\begin{figure}[htbp]
\includegraphics[width=0.35\textwidth]{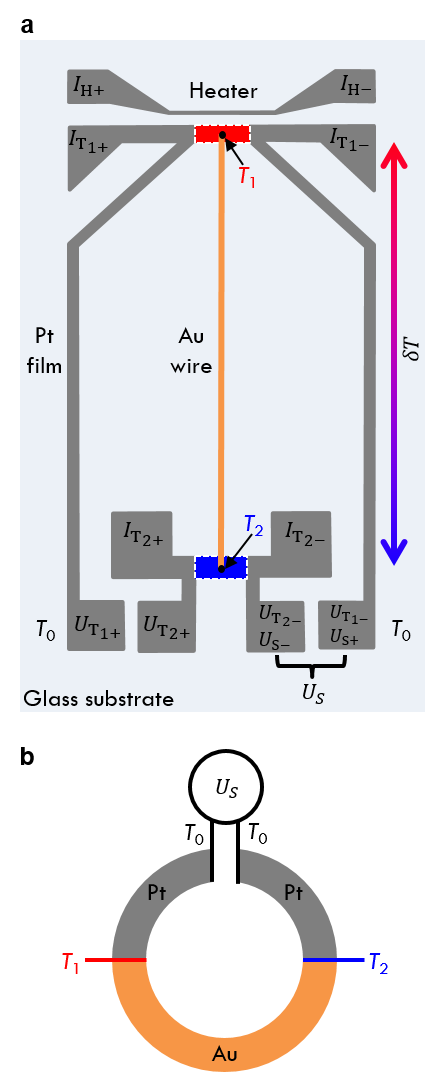}
\caption{{\bf Sketches of the measurement setup.} \textbf{a} Sketch of the thermoelectric micro lab. Platinum (Pt) was sputtered on a $\left(5\,\text{x}\,9\right)\,\text{mm}^{2}$ glass substrate. A line heater creates a temperature difference $\delta T = T_{1} - T_{2}$ along the sample by applying a current at the contacts $I_{\text{H+}}$ and $I_{\text{H-}}$. A gold (Au) wire $\left(\text{diameter}\,\,d=25\,\mu\text{m}\right)$ was bonded between the hot (red temperature $T_{1}$) and cold (blue temperature $T_{2}$) four-terminal resistance thermometers indicated by the corresponding current contacts ($I_{\text{T}_{1}},_{\text{T}_{2}}$) and voltage contacts ($U_{\text{T}_{1}},_{\text{T}_{2}}$). The thermovoltage $U_{\text{S}}$ was measured at the Pt pads with respect to the cold side of the sample. The Pt pads were kept at the same temperature $T_{0}$ in order to minimize parasitic thermovoltages. \textbf{b} Sketch of the measurement setup as a thermocouple consisting of gold (Au) and platinum (Pt). A temperature difference along both materials resulting from different junction temperatures ($T_{1}>T_{2}$) between Au and Pt produces a thermovoltage $U_{\text{S}}$.}   
\label{fig:Mikrolaborschema}
\end{figure}

A thermoelectric micro lab (TML) was designed with microlithography on a $5\,\text{mm}\,\text{x}\,9\,\text{mm}$ glass substrate and it is shown in figure \ref{fig:Mikrolaborschema} a. The TML involves a thermocouple, which consists of a thin sputtered platinum film (sputter target: $99,99\,\%$ platinum) and an attached bulk gold wire (diameter: $25\,\mu\text{m}$, purity: $99.99\,\%$). This gold wire creates a thermoelectric connection between the upper and lower part of the platinum film. A platinum line heater is used to generate a temperature difference $\delta T$ along the sample, which can be determined by four-terminal resistance thermometers. The resulting temperature difference along both materials due to the different junction temperatures ($T_{1}>T_{2}$) produces a thermovoltage $U_{\text{S}}$. The relative Seebeck coefficient between the gold wire and the thin platinum film with respect to the cold side is given by

\begin{equation}
S_{\text{Au,Pt}}=-\frac{U_{\text{S}_{\text{Au,Pt}}}}{\delta T}.
\end{equation}

Here, we prepared and investigated platinum films with a thickness of 134\,nm and 197\,nm. After the thermoelectric characterization of the samples in a flow cryostat in helium atmosphere at ambient pressure, a heat treatment was performed on the same samples. The heat treatment was carried out in a rapid thermal annealer in vacuum. The temperature of the annealer was gradually increased from $115\,^\circ\text{C}$ to $250\,^\circ\text{C}$ and finally to a maximum of $400\,^\circ\text{C}$. The temperature plateaus were held for two minutes each. A thermoelectric characterization of the same samples was performed after the heat treatment. For the X-ray investigations, $10\,\text{mm}\,\text{x}\,10\,\text{mm}$ large samples were prepared according to the same procedure with a thickness of $139$\,nm and $203$\,nm. X-ray experiments have been conducted with a lab-based diffractometer, with a Cu-$\text{K}_{\alpha}$ rotating anode source with a wavelength of $\lambda=0.154$\,nm. The thickness of all samples was determined by atomic force microscopy.

\begin{table}[htbp]
	\centering
		\begin{tabular}{|c|c|c|c|}
			\hline 
			Sample & $t\,(\text{nm})$ & $T_{\text{max}}\,(^\circ\text{C})$ & $RRR$ \\
			\hline 
			Pt 1 & $134 \pm 5$ & $-$ & $-$ \\
			Pt 2 & $134 \pm 5$ & $400$ & $8.45 \pm 0.01$ \\
			Pt 3 & $197 \pm 5$ & $-$ & $3.00 \pm 0.01$ \\
			Pt 4 & $197 \pm 5$ & $400$ & $8.26 \pm 0.01$ \\
			\hline
		\end{tabular}
		\caption{{\bf Thickness, heat treatment parameters and residual resistance ratio.} Overview of thickness $t$, maximum heat treatment temperature $T_{\text{max}}$ and residual resistance ratio $RRR$ of four thin platinum films. The thickness was determined by atomic force microscopy. The residual resistance ratio $RRR$ was determined as the ratio of the resistance at $290$\,K divided by the resistance of $20$\,K.}
		\label{tab:table1}
\end{table} 

\section{\label{sec:results}Results}

\subsection{X-ray}

\begin{figure}[htbp]
\includegraphics[width=0.5\textwidth]{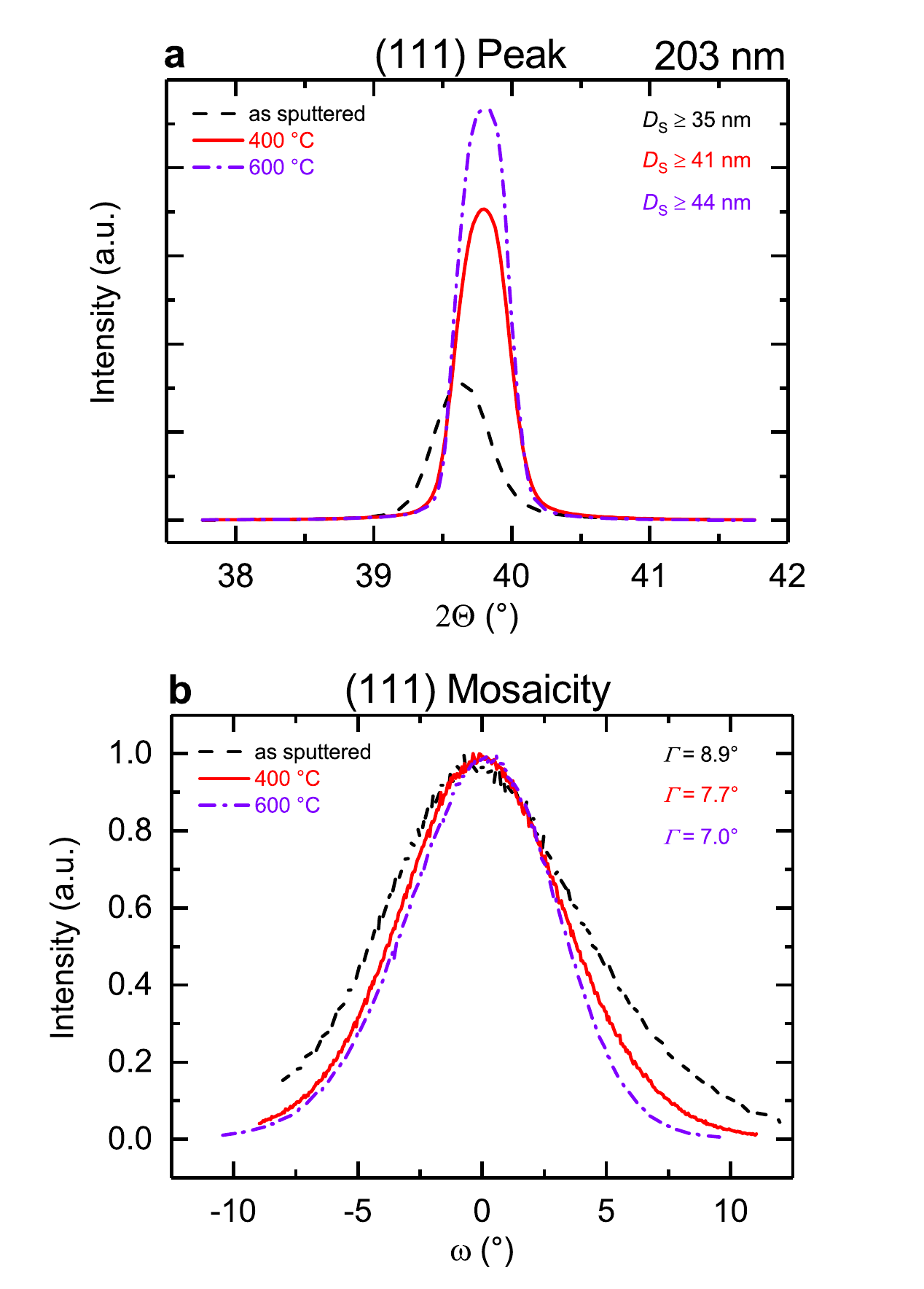}
\caption{{\bf X-ray diffractometry of a thin platinum film with a thickness of $203$\,nm as sputtered and with additional heat treatment.} \textbf{a} Intensity of the (111) Bragg reflection as a function of detector angle $2\Theta$ as sputtered, with heat treatment at a maximum temperature of $400\,^{\circ}\text{C}$ and $600\,^{\circ}\text{C}$. The average crystallographic grain size $D_{\text{S}}$ was determined by the Scherrer equation. \textbf{b} Intensity of the (111) Bragg peak as a function of the rocking angle variation $\omega$ as sputtered, with heat treatment at a maximum temperature of $400\,^{\circ}\text{C}$ and $600\,^{\circ}\text{C}$. The mosaicity $\Gamma$ of the platinum crystallites with a (111) texture is given by the full width at half maximum of the peak.}
\label{fig:Platinstrukturanalyse}
\end{figure}

X-ray measurements of polycrystalline platinum films with thicknesses of $139$\,nm and $203$\,nm exhibit four Bragg peaks corresponding to the platinum (111), (200), (220) and (311) reflections. The most intense Bragg peak of the $203$\,nm thin film is the (111) reflection (see figure \ref{fig:Platinstrukturanalyse} a, indicating a preferential orientation of the crystallites with a surface parallel (111) plane. Additional heat treatment and increased temperature of the heat treatment lead to an increase in the peak intensity and a shift of the peak position to larger detector angles, corresponding to smaller lattice constants. The position of the Bragg reflection at (111) of the annealed platinum films is in agreement with literature \cite{Wyckoff}. The average crystallographic grain size $D_{\text{S}}$ of the crystallites with a (111) orientation was estimated by the Scherrer equation \cite{Scherrer,Weber}

\begin{equation}
	D_{\text{S}}=\frac{K \lambda}{ \Delta(2\theta)\cos(\theta) }.
\end{equation}

$K$ is a dimensionless shape factor with a value of 0.9, $\lambda=0.154$\,nm is the X-ray wavelength. The broadening $\Delta(2\theta)$ is given by the full-width at half-maximum FWHM of the X-ray diffraction peak shown in figure \ref{fig:Platinstrukturanalyse} a and $\theta$ is the Bragg angle. $D_{\text{S}}$ of $139$\,nm and $203$\,nm from as sputtered thin platinum films are $D_{\text{S,139nm}}=(33\pm2)\,\text{nm}$ and $D_{\text{S,203nm}}=(35\pm2)\,\text{nm}$, respectively. This size increases with heat treatment at $400\,^\circ\text{C}$ to $D_{\text{S,139nm,400C}}=(41\pm1)\,\text{nm}$ and $D_{\text{S,203nm,400C}}=(41\pm1)\,\text{nm}$, respectively. 

Furthermore, the mosaicity $\Gamma$, which is a measure of the spread of crystal plane orientations, was determined for the (111) Bragg reflection. Figure \ref{fig:Platinstrukturanalyse} b shows a rocking scan where the sample angle $\omega$ is varied for a fixed detector angle $2\theta$. As the (111) lattice planes are not all perfectly parallel to the substrate surface, intensity is found within an angular distribution of FWHM. $\Gamma$ of $139$\,nm and $203$\,nm as sputtered thin platinum films are $\Gamma_{\text{139nm}}=(11.0\pm0.4)\,^\circ$ and $\Gamma_{\text{203nm}}=(8.9\pm0.2)\,^\circ$, respectively. Thicker platinum films therefore have a more perfect texture corresponding to a narrower distribution of crystallite tilt angles. The tilt decreases and so the crystal quality increases with heat treatment at $400\,^\circ\text{C}$ to $\Gamma_{\text{139nm,400C}}=(9.5\pm0.2)\,^\circ$ and $\Gamma_{\text{203nm,400C}}=(7.7\pm0.2)\,^\circ$, respectively. 

\subsection{Electrical measurements}

\begin{widetext}

\begin{figure}[htbp]
\includegraphics[width=1.0\textwidth]{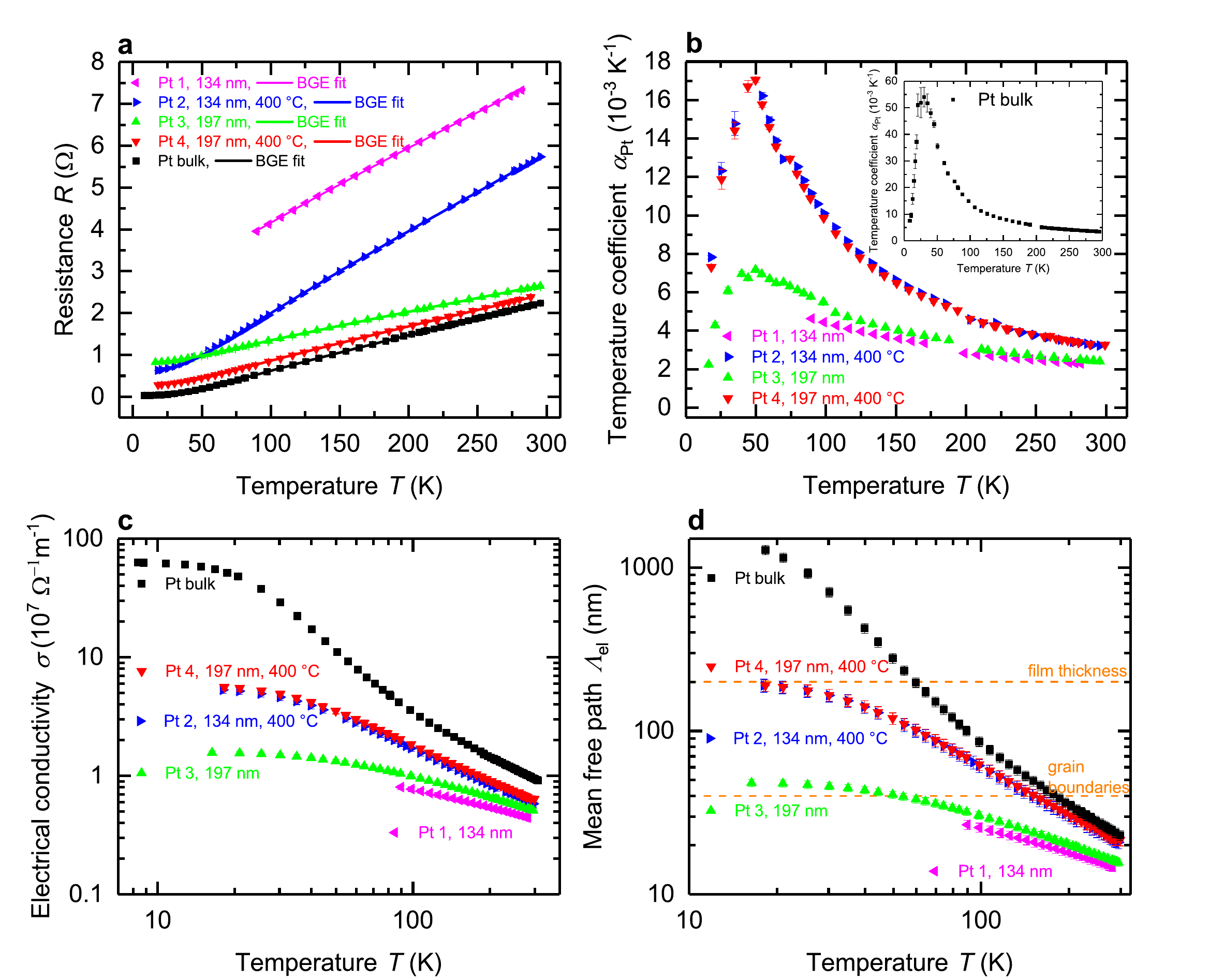}
\caption{{\bf Electrical measurements of thin platinum films as-sputtered and with heat treatment and of the bulk.} \textbf{a} Four-terminal resistance $R$ as a function of the bath temperature $T$. The Bloch-Grüneisen equation (BGE) was fitted to the data in order to determine the temperature dependence of the resistance and to calculate the Debye temperature. \textbf{b} Temperature coefficient $\alpha_{\text{Pt}}$ of the resistance as a function of the bath temperature $T$. The inset shows temperature coefficient of the bulk material. \textbf{c} Electrical conductivity $\sigma$ as a function of the bath temperature $T$. \textbf{d} Mean free path of the electrons $\Lambda_{\text{el}}$ as a function of the bath temperature $T$. The mean free path of as sputtered thin platinum films is mainly limited by grain boundaries at low temperatures. The mean free path of thin platinum films with heat treatment is mainly limited by the film thickness.}
\label{fig:Platin_Elektrische_Messungen_Alternative}
\end{figure}

\end{widetext}

The Bloch-Grüneisen equation was used to fit the temperature-dependent resistance of the platinum films, which show the expected metallic behavior given in figure \ref{fig:Platin_Elektrische_Messungen_Alternative} a, in order to determine the Debye temperature $\Theta_{\text{D}}$ of the material. All platinum films with heat treatment and the bulk material (wire diameter $d=25\,\mu\text{m}$) agree with the literature value of $\Theta_{\text{D}}=240$\,K\,\,\,\,\cite{Stojanovic}. Pt 3 ($197$\,nm, without heat treatment) has a reduced Debye temperature $\Theta_{\text{D,Pt 3}}=191$\,K compared to the literature. This can be attributed to the microstructure. 

Furthermore, the residual resistance ratio $RRR$ was determined as the ratio of the resistance at $290$\,K divided by the resistance of $20$\,K in order to compare the quality of the sputtered platinum films and the influence of the heat treatment. Bulk platinum has the highest residual resistance ratio $RRR_{\text{bulk}}=50.3\pm0.1$. Thin films have a reduced residual resistance ratio compared to the bulk given in table \ref{tab:table1}. 
 
Figure \ref{fig:Platin_Elektrische_Messungen_Alternative} b shows the temperature coefficient $\alpha_\text{Pt}$ of the resistance, which depends on the thickness of the platinum films. Larger film thickness and additional heat treatment lead to an increase of the temperature coefficient. Like the residual resistance ratios, the temperature coefficients of Pt 2 ($134$\,nm, with heat treatment) and Pt 4 ($197$\,nm, with heat treatment) are similar. The four-terminal resistance of the cold thermometer was used to determine the electrical conductivity $\sigma$ of the thin platinum films shown in figure \ref{fig:Platin_Elektrische_Messungen_Alternative} c. The electrical conductivity of the platinum bulk wire $\sigma_{\text{Pt,bulk}}$ is larger than $\sigma_{\text{Pt,film}}$. $\sigma_\text{Pt,film}$ depends on the film thickness, can be increased by heat treatment and reaches a maximum at low bath temperatures. 

Figure \ref{fig:Platin_Elektrische_Messungen_Alternative} d features the electron mean free path $\Lambda_{\text{el}}$ of the of the thin films and of the bulk. The electron mean free path of the thin films is reduced compared to the bulk and increases with increasing film thickness and can be further increased by heat treatment. At low bath temperatures, the mean free path reaches a maximum limited by the film thickness and the structural properties.

\subsection{Seebeck measurements}

\begin{figure}[htbp]
\includegraphics[width=0.5\textwidth]{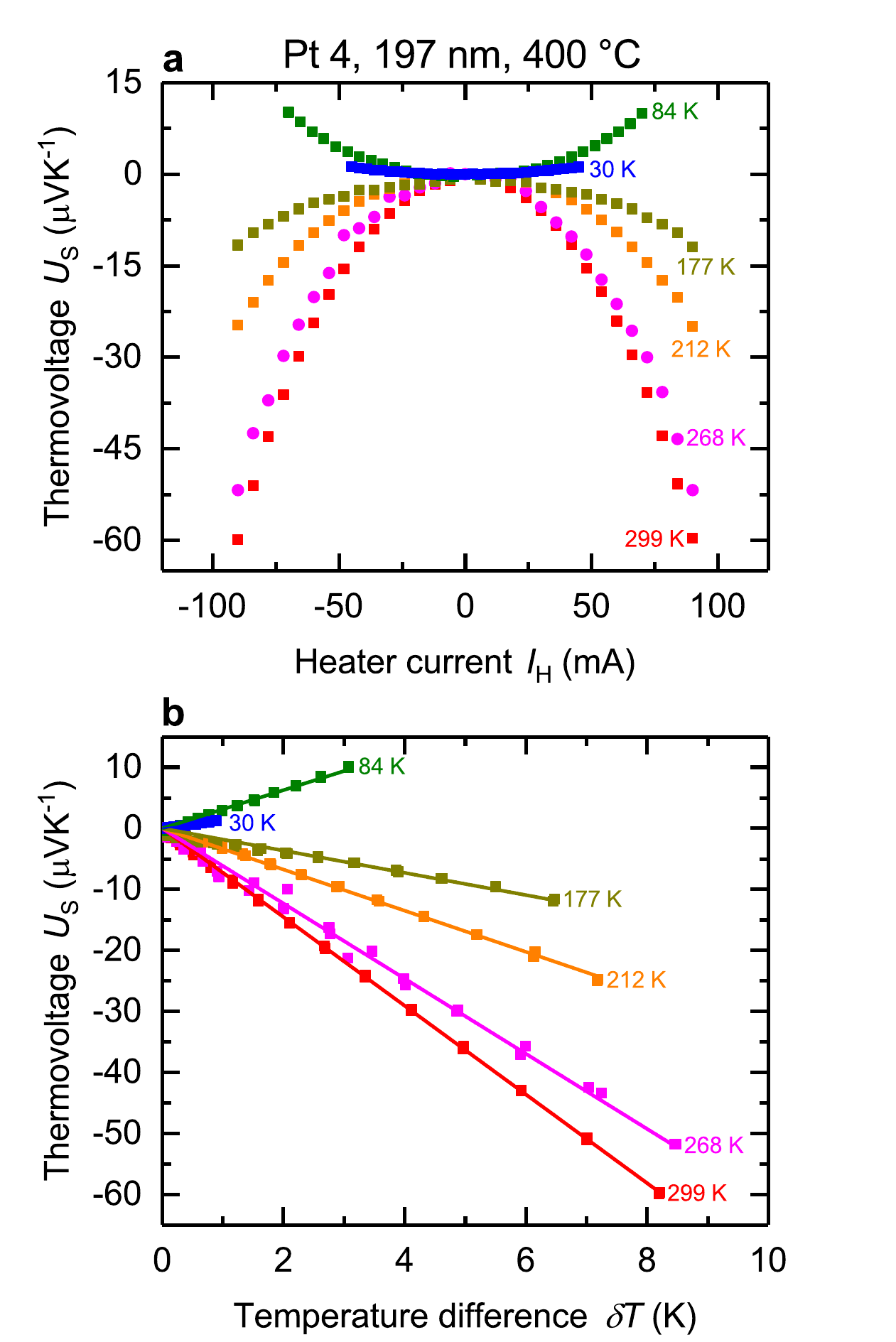}
\caption{{\bf Thermovoltage as a function of the heater current and temperature difference of Pt 4.} \textbf{a} Thermovoltage $U_{\text{S}}$ as a function of heater current $I_{\text{H}}$ at different bath temperatures $T$. The sign of the thermovoltage changes below $T\approx 150$\,K. \textbf{b} Thermovoltage $U_{\text{S}}$ as a function of temperature difference $\delta T$ at different bath temperatures $T$. The slope of the fitted solid lines gives the relative Seebeck coefficient between the thin platinum films and the bulk gold wire.}
\label{fig:Platin_Seebeck_Plattform_Au_Pt_197nm_400C_Thermospannung}
\end{figure}

In addition to the electrical characterization, the temperature-dependent Seebeck coefficient of thin platinum films and bulk platinum relative to a bulk gold wire (wire diameter $d=25\,\mu\text{m}$) was determined by measuring the thermovoltage $U_{\text{S}}$ as a function of the heater current of the line heater. Figure \ref{fig:Platin_Seebeck_Plattform_Au_Pt_197nm_400C_Thermospannung} shows the parabolic behavior of the thermovoltage indicating the thermoelectric effect. The temperature difference $\delta T$ was determined by four-terminal resistance measurements of the thermometers at the hot and cold side of the measurement platform. The slope of the function $U_{\text{S}}(\delta T)$ gives the relative Seebeck coefficient

\begin{equation}
	S_{\text{Au,Pt}}=S_{\text{Au}}-S_{\text{Pt}}=-\frac{U_{\text{S}_{\text{Au,Pt}}}}{\delta T}.
\end{equation}

\begin{figure}[htbp]
\includegraphics[width=0.5\textwidth]{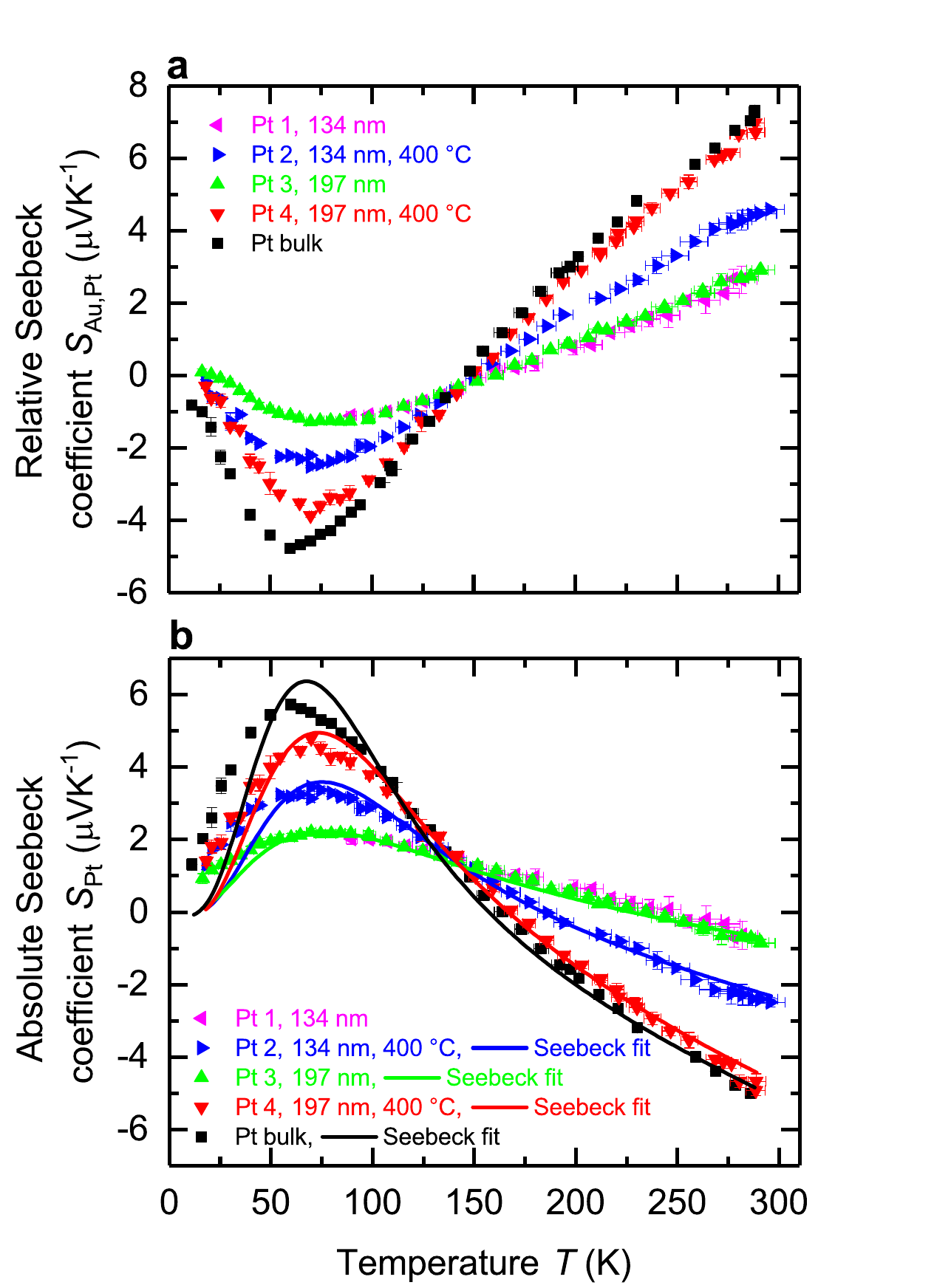}
\caption{{\bf Relative and absolute Seebeck coefficient of thin platinum films as sputtered and with heat treatment and bulk material.} \textbf{a} Relative Seebeck coefficient $S_{\text{Au,Pt}}$ of platinum relative to a gold wire with a diameter of $25\,\mu\text{m}$ as a function of the bath temperature $T$. \textbf{b} Absolute Seebeck coefficient $S_{\text{Pt}}$ of platinum as a function of the bath temperature $T$. The corresponding solid lines represent the Seebeck fit of the data. Absolute Seebeck coefficient of bulk gold $S_{\text{Au}}$ was taken from \cite{HuebenerAu2}.}
\label{fig:Platin_Seebeck_Messungen}
\end{figure}

The relative Seebeck coefficient of the thin films and the bulk material decreases with decreasing bath temperature as depicted in figure \ref{fig:Platin_Seebeck_Messungen} a. In order to determine the absolute Seebeck coefficient of platinum $S_{\text{Pt}}$, the absolute Seebeck coefficient of bulk gold $S_{\text{Au}}$ was taken from \cite{HuebenerAu2}. $S_{\text{Pt}}$ is given in figure \ref{fig:Platin_Seebeck_Messungen} b.

\section{\label{sec:discussion}Discussion}

\subsection{Structural and electrical characterization}

X-ray diffraction analysis gives the average crystallographic grain size $D_{\text{S}}$, which was determined by the Scherrer equation \cite{Scherrer,Weber} from the (111) Bragg reflections of figure \ref{fig:Platinstrukturanalyse} a. The determined values indicate only a lower limit for the crystallographic grain size. The actual grain size can be larger. Furthermore, the X-ray measurements only provide information about the grain size in the growth direction. The in-plane grain size can be significantly larger than the grain size in the growth direction. In addition to the average crystallographic grain size, the morphological grain size is discussed in literature \cite{Salvadori,Melo}. It was proposed that the morphological grains, which were determined by scanning tunneling microscopy, are agglomerates of crystallographic grains and that the morphological grain size is increasing with increasing film thickness much further than the crystallographic grain size \cite{Salvadori,Melo}. 

The structural properties like film thickness and grain size can be linked with the mean free path of the electrons in order to investigate the transport properties of the thin platinum films. The electron mean free path $\Lambda_{\text{el}}$ was determined from the electrical conductivity of the thin films with Matthiessen rule \cite{Kojda2}.

\begin{equation}
	\Lambda_{\text{el,f}}(T)^{-1} = \Lambda_{\text{el,b}}(T)^{-1} + \Lambda_{\text{el,s,gb}}^{-1}
\end{equation}

is the inverse thin film electron mean free path, $\Lambda_{\text{el,b}}(T)^{-1}$ is the inverse mean free path of the bulk and $\Lambda_{\text{el,s,gb}}^{-1}$ is the inverse temperature-independent scattering length of the electrons due to surface and grain boundary scattering. 

\begin{equation}	
	\Lambda_{\text{el,f}}(T) = \frac{\sigma_{\text{f}}(T)}{\sigma_{\text{f}}(\text{RT})} \Lambda_{\text{el,f}}(\text{RT})
\end{equation}
and
\begin{equation}
	\Lambda_{\text{el,b}}(T) = \frac{\sigma_{\text{b}}(T)}{\sigma_{\text{b}}(\text{RT})} \Lambda_{\text{el,b}}(\text{RT})
\end{equation}	

are the electron mean free paths of the thin films and of the bulk material, respectively. $\Lambda_{\text{el,b}}(\text{RT}) = 23\,\text{nm}$ is given in literature at room temperature \cite{Stojanovic}. Figure \ref{fig:Platin_Elektrische_Messungen_Alternative} d shows the electron mean free path as a function of the bath temperature. Table \ref{tab:table2} shows the electron mean free path at room temperature and the temperature-independent scattering length of the electrons. Pt 3 ($197$\,nm, without heat treatment) has a mean free path which saturates at low bath temperatures at $\Lambda_{\text{el,s,gb}}=(48\pm1)\,\text{nm}$. Pt 2 ($134$\,nm, with heat treatment) and Pt 4 ($197$\,nm, with heat treatment) have nearly the same temperature dependence of the electron mean free path, which is in the order of magnitude of the film thickness. Heat treatment increases the mean free path. From the structural properties, we can conclude that the electron mean free path at low bath temperatures is mainly limited by grain boundaries for thin films without heat treatment. For thin films with heat treatment, the main limitation is set by the film thickness.

\begin{table}[htbp]
	\centering
		\begin{tabular}{|c|c|c|c|c|}
			\hline 
			Sample & $t$\,(nm) & $T_{\text{max}}\,(^\circ\text{C})$ & $\Lambda_{\text{el,f}}(\text{RT})$\,(nm) & $\Lambda_{\text{el,s,gb}}$\,(nm) \\
			\hline 
			Pt 1 & $134$ & - & $14\pm1$ & $36\pm1$ \\
			Pt 2 & $134$ & $400$ & $21\pm2$ & $212\pm2$ \\
			Pt 3 & $197$ & - & $16\pm1$ & $48\pm1$ \\
			Pt 4 & $197$ & $400$ & $22\pm2$ & $219\pm2$ \\
			\hline
		\end{tabular}
		\caption{{\bf Electron mean free path.} Mean free path of the samples with thickness $t$ and maximum heat treatment temperature $T_{\text{max}}$. $\Lambda_{\text{el,f}}(\text{RT})$ is the electron mean free path of the thin films at room temperature and $\Lambda_{\text{el,s,gb}}$ is the temperature-independent mean free path of the electrons due to surface and grain boundary scattering. $\Lambda_{\text{el,b}}(\text{RT}) = 23\,\text{nm}$ of bulk is given in literature at room temperature \cite{Stojanovic}. }
		\label{tab:table2}
\end{table}

\subsection{Seebeck coefficient}

In general, the Seebeck coefficient is the sum of two parts. The thermodiffusion part and the phonon drag part \cite{MacDonald,HuebenerPt1,HuebenerPt2}:
\begin{equation}
	S = S_{\text{diff}} + S_{\text{ph}}.
	\label{eq:Seebeckallgemein}
\end{equation}
$S_{\text{diff}}$ is the contribution due to the thermodiffusion of the charge carriers as described by Mott's formula \cite{CutlerMott,HuebenerPt1,HuebenerPt2}:
\begin{equation}
	S_{\text{diff}} = \frac{\pi^{2} k^{2}_{\text{B}} T}{3 e} \left(\frac{\partial \ln \sigma(\epsilon)}{\partial \epsilon} \right) \bigg|_{\epsilon_{F}}
	\label{eq:MottFormel}
\end{equation}
with $k_{\text{B}}$ is the Boltzmann constant, $T$ is the bath temperature, $e$ is the elementary charge and $\frac{\partial \ln\sigma(\epsilon)}{\partial\epsilon}\big|_{\epsilon_{F}}$ is the derivative of the electrical conductivity according to the energy at the Fermi energy. 

$S_{\text{ph}}$ is the contribution due to the phonon drag effect. The phonon drag effect is based on the electron-phonon interaction. A momentum transfer from phonons to electrons leads to an increase of the Seebeck coefficient \cite{MacDonald}. The phonon drag part of the Seebeck coefficient is connected with the specific heat of the phonons \cite{MacDonald,HuebenerPt2}
\begin{equation}
	C_{\text{ph} (T)} = 9 n k_{\text{B}} \left(\frac{T}{\Theta_{\text{D}}}\right)^{3} \int_{0}^{\frac{\theta_{\text{D}}}{T}} \frac{x^{4} \exp (x)}{\left( \exp \left( x \right) - 1\right)^{2}} \text{d}x
	\label{eq:SpezifischeWaerme}
\end{equation}
with the number of charge carriers per volume $n$ and $x = \frac{\hbar \omega_{\text{D}}}{k_{\text{B}} T} = \frac{\Theta_{\text{D}}}{T}$ which is given by the reduced Planck constant $\hbar$ and by the Debye frequency $\omega_{\text{D}}$.

There are low and high temperature approaches to describe the phonon drag part of the Seebeck coefficient \cite{MacDonald}. The low temperature approach can be described by $S_{\text{ph}}=\frac{C_{\text{ph,low}}}{3ne}$. According to the Debye model the specific heat $C_{\text{ph,low}}$ goes with $T\rightarrow 0$ to $C_{\text{ph,low}} \propto T^{3}$ and for this reason $S_{\text{ph,low}} \propto T^{3}$ \,\, \cite{MacDonald}. The resulting Seebeck coefficient can be written as $S_{\text{low}}=S_{\text{diff}}+S_{\text{ph,low}} = a T + b T^{3}$. $a$ and $b$ are variables.

The high temperature approach has to be modified compared to the low temperature approach because the phonon-phonon scattering time $\tau_{\text{pp}}$ and the scattering time of the electron-phonon interaction $\tau_{\text{ep}}$ has to be taken into account \cite{MacDonald,HuebenerPt2}. The Debye model predicts a constant specific heat $\left( C_{\text{ph,high}} \approx 3 n k_{\text{B}}\right)$ for temperatures above the Debye temperature $\Theta_{\text{D}}$ and due to phonon-phonon Umklapp scattering $S_{\text{ph,high}} \propto \frac{1}{T}$ \,\, \cite{MacDonald}. This results in $S_{\text{high}}=S_{\text{diff}}+S_{\text{ph,high}} = c T + d\frac{1}{T}$. $c$ and $d$ are variables. 

The combination of the low temperature approach, the high temperature approach and additionally a characterization of the intermediate regime gives 
\begin{equation}
	S_{\text{ph}} = \frac{C_{\text{ph}} (T) }{3ne} \gamma = \frac{C_{\text{ph}} (T) }{3ne} \frac{\tau_{\text{pp}}}{\tau_{\text{pp}}+\tau_{\text{ep}}}\,\,\,\,\text{\cite{MacDonald,HuebenerPt2}},
	\label{eq:SeebeckPhononenDrag1}
\end{equation}
hence
\begin{equation}
	S_{\text{ph}} = \frac{C_{\text{ph}} (T) }{3ne} \frac{1}{1 + \frac{\tau_{\text{ep}}}{\tau_{\text{pp}}}} = \frac{C_{\text{ph}} (T) }{3ne} \frac{1}{1 + F_{\tau} T \exp\left(-\frac{\Theta_{\text{D}}}{T}\right)}.
	\label{eq:SeebeckPhononenDrag2}
\end{equation}
The relation $\frac{\tau_{\text{ep}}}{\tau_{\text{pp}}} = F_{\tau} T \exp\left(-\frac{\Theta_{\text{D}}}{T} \right)$ was suggested by and adapted from \cite{HuebenerPt2,Walker}. This equation assumes that the phonon-phonon interaction becomes dominant at high temperatures \cite{HuebenerPt2,Walker}. This results in the $\gamma$-factor
\begin{equation}
	\gamma = \frac{1}{1 + \frac{\tau_{\text{ep}}}{\tau_{\text{pp}}}} = \frac{1}{1 + F_{\tau} T \exp\left(-\frac{\Theta_{\text{D}}}{T}\right)}.
	\label{eq:GammaFaktor}
\end{equation}
Finally, we combine the approach of the phonon drag part with the specific heat of the phonons and the thermodiffusion part $S_{\text{diff}} = F_{\text{diff}} \frac{T}{\Theta_{D}}$ to the formula, which describes the Seebeck coefficient over a wide temperature range
\begin{equation}
	S = F_{\text{diff}} \frac{T}{\Theta_{\text{D}}} + \frac{F_{\text{ph}} \left(\frac{T}{\Theta_{\text{D}}}\right)^{3} \int_{0}^{\frac{\theta_{\text{D}}}{T}} \frac{x^{4} \exp (x)}{\left( \exp \left( x \right) - 1\right)^{2}} \text{d}x}{1 + F_{\tau} T \exp\left(-\frac{\Theta_{\text{D}}}{T}\right)}.
	\label{eq:SeebeckKomplett}
\end{equation}

The Seebeck fit and measurement results are given in figure \ref{fig:Platin_Seebeck_Messungen} b. 
The phonon drag peak appears between bath temperatures of $65$\,K and $75$\,K. A possible reason for the deviation of the Seebeck fit from the measurement results below $50$\,K is that the phonon drag part is very sensitive to impurities \cite{Bailyn2,HuebenerPt1,HuebenerAu2}. The Seebeck fit provides different parameters, which are given in table \ref{tab:table3}.

The thermodiffusion part is described by the fit parameter $F_{\text{diff}}$. The platinum films without heat treatment have the smallest thermodiffusion contribution to the absolute Seebeck coefficient, which can be attributed to an increased scattering rate at grain boundaries. Pt 2 ($134$\,nm, with heat treatment) and Pt 4 ($197$\,nm, with heat treatment) exhibit a film thickness dependence of the thermodiffusion part on the bath temperature. The fit parameter of the thermodiffusion part of Pt 2 is $F_{\text{diff,Pt 2}}=(-2.6\pm0.1)\,\mu \text{VK}^{-1}$ and of Pt 4 is $F_{\text{diff,Pt 4}}=(-4.7\pm0.1)\,\mu \text{VK}^{-1}$. This difference can be explained by size effects like surface scattering, which is more likely to happen in Pt 2 than in Pt 4. The thermodiffusion part of Pt 4 and bulk have nearly the same temperature dependence and magnitude and it seems that Pt 4 reaches the upper limit of the thermodiffusion part, which is provided by the bulk material.

Compared to the thermodiffusion part, the phonon drag part vanishes at high bath temperatures. $F_{\text{ph}}$ estimates the strength of the phonon drag on the absolute Seebeck coefficient. Pt 3 ($197$\,nm, without heat treatment) has the smallest value of the fit parameter $F_{\text{ph,Pt 3}}=(24\pm2)\,\mu \text{VK}^{-1}$. This value increases with increasing film thickness and with additional heat treatment and reaches its maximum for the bulk material with $F_{\text{ph,bulk}}=(47\pm2)\,\mu \text{VK}^{-1}$. 

The ratio $|F_{\text{ph}}$/$F_{\text{diff}}|$ of bulk and of the thin films is approximately $10$. Except for Pt 4 ($197$\,nm, with heat treatment), which is slightly lower. This ratio indicates that the thermodiffusion and phonon drag part are reduced by nearly the same factor, when the film thickness is decreasing. To further illustrate that the thermodiffusion and the phonon drag part are related to each other, we introduce $F_{\tau}$, which gives the ratio of the scattering time of the electron-phonon and phonon-phonon interaction and determines the $\gamma$-factor. The $\gamma$-factor, see equation \ref{eq:GammaFaktor}, is a number between 0 and 1, which depends on the interaction between phonons and electrons. For $T \ll \Theta_{\text{D}}$, $\gamma \approx 1$, means electron-phonon interaction is dominant compared to phonon-phonon interaction. Phonon-phonon interaction is dominant compared to electron-phonon for $\gamma \approx 0$. 

\begin{figure}[htbp]
\includegraphics[width=0.5\textwidth]{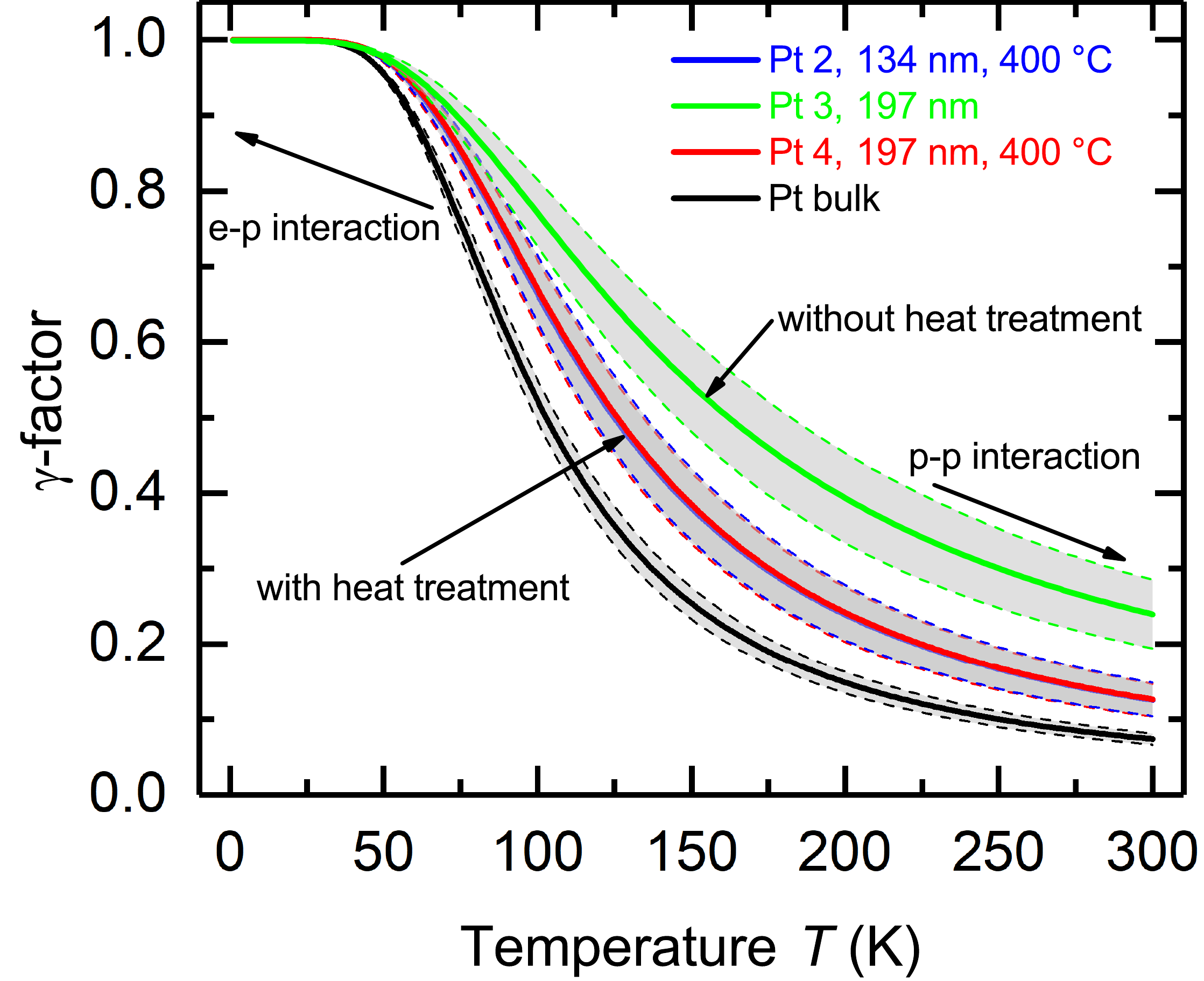}
\caption{\label{fig:Platin_gamma-Faktor} The panel shows the $\gamma$-factor of thin platinum films as sputtered and with heat treatment and bulk material as a function of bath temperature. The $\gamma$-factor is determined by the electron-phonon (e-p) and phonon-phonon (p-p) interaction. The gray shaded area around the solid lines marks the uncertainty. $\gamma=1$ means that the electron-phonon interaction compared to the phonon-phonon interaction is dominant. $\gamma=0$ means that the phonon-phonon interaction compared to the electron-phonon interaction is dominant. The curves of Pt 2 and Pt 4 have a similar temperature dependence and are lying on top of each other.}
\end{figure}

The $\gamma$-factor as a function of temperature for thin films and bulk is given in figure \ref{fig:Platin_gamma-Faktor}. For all temperatures applies: $\gamma_{\text{film}}>\gamma_{\text{bulk}}$. 
This means that there is an increased amount of electron-phonon interaction compared to phonon-phonon interaction in the thin films than in the bulk. The influence of the phonon drag part on the absolute Seebeck coefficient dominates in thin films compared to the bulk. For example, the thermodiffusion part of Pt 3 is reduced by 77\,\% towards bulk, resulting in a significant effect of the phonon drag part even at room temperature. This difference can be explained by the inner and outer interfaces of the the thin films and the resulting grain boundary scattering. 

Decreasing temperatures lead to an increase of $\gamma_{\text{film}}$ and $\gamma_{\text{bulk}}$. This indicates that the electron-phonon interaction is becoming more dominant compared to phonon-phonon interaction. At temperatures below 50\,K, the thermodiffusion part tends to 0 and the $\gamma$-factor tends to 1. The reason for this behavior can be attributed to the phonon-phonon interaction, which is negligible compared to the electron-phonon interaction. 

Pt 2 ($134$\,nm, with heat treatment) and Pt 4 ($197$\,nm, with heat treatment) have the same $\gamma$-factor but different absolute Seebeck coefficients, indicating that the ratio of electron-phonon interaction compared to the phonon-phonon interaction is the same but the absolute amount of electron-phonon and phonon-phonon interaction is larger in thicker platinum films with heat treatment, because the essential limitation is no longer caused by grain boundaries, but by the film thickness.

\begin{widetext}

\begin{table}[htbp]
	\centering
		\begin{tabular}{|c|c|c|c|c|c|c|c|}
			\hline 
			Sample & $t$\,(nm) & $T_{\text{max}}\,(^\circ\text{C})$ & $F_{\text{ph}}\,(\mu \text{VK}^{-1})$  & $F_{\text{$\tau$}}\,(\text{K}^{-1})$ & $F_{\text{diff}}\,(\mu \text{VK}^{-1})$ & $|F_{\text{ph}}$/$F_{\text{diff}}|$ & $S_{\text{Pt}}\,(\mu \text{VK}^{-1})$ at $280\,$K \\
			\hline 
			Pt 1 & 134 & - & - & - & - & - & $-0.6 \pm 0.4$ \\
			Pt 2 & 134 & 400 & $24 \pm 2$ & $0.05 \pm 0.01$ & $-2.6 \pm 0.1$ & $9.2 \pm 0.8$ & $-2.3 \pm 0.3$ \\
			Pt 3 & 197 & - & $12 \pm 1$ & $0.020 \pm 0.005$ & $-1.1 \pm 0.1$ & $10.9 \pm 1.3$ & $-0.7 \pm 0.1$ \\
			Pt 4 & 197 & 400 & $35 \pm 2$ & $0.05 \pm 0.01$ & $-4.7 \pm 0.1$ & $7.4 \pm 0.5$ & $-4.7 \pm 0.2$ \\
			Bulk & - & - & $47 \pm 2$ & $0.09 \pm 0.01$ & $-4.8 \pm 0.1$ & $9.8 \pm 0.5$ & $-4.8 \pm 0.1$ \\
			\hline
		\end{tabular}
		\caption{{\bf Seebeck fit parameters.} Parameters of Seebeck fit for each sample with thickness $t$ and maximum heat treatment temperature $T_{\text{max}}$. $F_{\text{ph}}$ is the parameter which describes the intensity of the phonon drag part. $F_{\text{$\tau$}}$ gives the ratio of the scattering time of the electron-phonon interaction and of the phonon-phonon interaction. $F_{\text{diff}}$ describes the intensity of the thermodiffusion part. $|F_{\text{ph}}$/$F_{\text{diff}}|$ gives the modulus ratio between the fit parameter of the thermodiffusion part and phonon drag part. $S_{\text{Pt}}\,(\mu \text{VK}^{-1})$ is the absolute Seebeck coefficient of platinum at $280\,$K. Due to the lack of low temperature data, the fit parameters of Pt 1 ($197$\,nm, without heat treatment) are not given. }
		\label{tab:table3}
\end{table} 

\end{widetext}

\section{\label{sec:conclusion}Conclusion}

In this work, we performed thermoelectric and structural characterizations of thin sputtered platinum films. The influence of heat treatment and film thickness on the electrical conductivity and the absolute Seebeck coefficient were investigated. Additional heat treatment and a larger film thickness increase the crystal quality of sputtered platinum films. The electrical conductivity and the absolute Seebeck coefficient are reduced compared to the bulk due to size effects like surface and boundary scattering. We find that structural properties like grain size and film thickness, which limit the electron mean free path, influence the absolute Seebeck coefficient. For the phonon drag part of the absolute Seebeck coefficient, the electron-phonon interaction compared to the phonon-phonon interaction plays a more dominant role in thin films than in bulk. If the mean free path of thin metallic films is in the order of the film thickness, the absolute Seebeck coefficient of bulk is no more valid. This has to be taken into account, when using thin platinum films as a reference material for the determination of the absolute Seebeck coefficient. Due to the influence of the microstructure, metallic interconnects can be tailored in a way, that the relative Seebeck coefficient can be reduced to zero, which is interesting for low-noise applications.

\section{\label{sec:acknowledgements}Acknowledgements}

We acknowledge partial financial support by the German Science Foundation (DFG-SPP1386).

\end{document}